\documentclass[%
aip,
 amsmath,amssymb,
 reprint,%
]{revtex4-1}
\usepackage{graphicx}
\usepackage{dcolumn}
\usepackage{bm}

\usepackage[utf8]{inputenc}
\usepackage[T1]{fontenc}
\usepackage{mathptmx}
\usepackage{etoolbox}

\renewcommand{\d}{\mathrm{d}}

\begin{document}


\title{Microphysical boundary condition for the electron kinetics of a plasma} 



\author{Felix Willert}

\email[]{felix.willert@uni-greifswald.de}
\affiliation{Institut für Physik, Universität Greifswald, Felix-Hausdorff-Str. 6, 17489 Greifswald, Germany}

\author{Clemens Hoyer}

\email[]{clemens.hoyer@stud.uni-greifswald.de}
\affiliation{Institut für Physik, Universität Greifswald, Felix-Hausdorff-Str. 6, 17489 Greifswald, Germany}

\author{Gordon K. Grubert}
\affiliation{Universitätsrechenzentrum, Universität Greifswald, Felix-Hausdorff-Str. 18, 17489 Greifswald, Germany}

\author{Franz X. Bronold}

\email[]{bronold@uni-greifswald.de}
\affiliation{Institut für Physik, Universität Greifswald, Felix-Hausdorff-Str. 6, 17489 Greifswald, Germany}

\date{\today}

\begin{abstract}
We derive and implement a suitable boundary condition for the kinetic description of the electrons inside a plasma, which takes into account microphysical processes inside the wall. It is based on the surface scattering kernel, which describes the scattering cascade of the electron in the solid and the excitation of secondary electrons. The resulting boundary condition is inelastic, angle- and energy-dependent. The implementation for a Boltzmann equation solved by a Legendre polynomial expansion method is presented, elucidating the modest additional computational cost of the new boundary condition. Results, indicating the influence of the inelasticity, are shown for the example of a silicon wall facing argon, helium and oxygen plasmas, but the described construction is also valid for other materials. An effective reflection coefficient is defined to compare the results with previously used boundary conditions.
\end{abstract}


\maketitle 

\section{Introduction}
As essentially all human-made plasmas are bounded by condensed matter, boundary conditions which describe the interaction of the electrons with the containing wall properly are crucial for a realistic theoretical description and simulation of gas discharges. To obtain a physically meaningful boundary condition and experimentally reproducible results one has to relate macroscopic quantities and thereby measurable properties to distribution functions at the margins of the plasma \cite{Becker2010}.

An important feature that the boundary condition has to describe is the occurrence of secondary electrons. These secondary electrons can arise from heavy species impinging the wall \cite{Phelps_1999}, where they can play an important role in breakdown phenomena \cite{Breakdown2007,Braginsky_2012}. But also electron induced secondary electrons can occur, which are of importance in many types of gas discharges, e. g. plasma thrusters \cite{Gascon2003,Sydorenko2006,Taccogna2009} and capacitively coupled plasmas for industrial applications like sputtering, etching, and plasma immersed ion implantation \cite{Horvath_2017}. Apart from that, they are essential for scanning electron microscopes \cite{Seiler1983,JOY1996}.

Although secondary electrons are not always regarded in numerical simulations and instead a perfect absorber for electrons is assumed, a variety of works \cite{Taccogna2004,Bradshaw_2024,Cichocki_2023,CAGAS2020} made progress in the inclusion of more realistic descriptions of the bounding solid in plasma simulations. There is a variety of phenomenological models \cite{PhysRevSTAB.5.124404,Horvath_2017,Vaughan,Dionne1973} that are commonly used to describe the interaction of an electron with a solid. The complexity of these models reaches from simple constants to a set of equations with more than twenty parameters. These parameters usually have to be fitted to experimental data, but due to a lack of adequate data often many parameters have to be guessed (most of the time the data are only available for some, rather high energies and complete emission spectra are, if at all, only measured for a single impact energy and a single incident angle). Monte Carlo simulations \cite{Ding1996} were also used for the description of plasma-solid-interaction.

Compared to phenomenological models, we are concerned with a theoretical and microphysical one, which was presented in one of our previous works \cite{PhysRevE.110.035207}. The model describes the scattering cascade of an electron, which impinges a solid, as well as impact ionization, which leads to the creation of secondaries. It uses a rough approximation for the band structure of the solid and accounts for electron-phonon, electron-ion-core and electron-electron interaction. This model leads to the so-called \textit{Surface Scattering Kernel} (SSK), which links the distribution function of the electrons leaving the plasma and entering the wall to the distribution function of the electrons reentering the plasma. A drawback of this approach is that the SSK cannot be expressed analytically and the description of the solid is only valid up to a certain energy, above which the data for the SSK has to be extrapolated to higher energies.

Therefore, the main goal of this article is to present a new boundary condition based on the SSK that can be used for solving a Boltzmann transport equation in a numerically cost efficient way. The construction of the boundary condition is carried out with data for the SSK for Si, but it works with other materials as well, provided the SSK has been calculated for them. Furthermore, it will be shown that the new boundary condition produces relevant changes in the bulk and the sheath compared to simpler, previously used, material insensitive boundary conditions. The different magnitude of these changes in different gases are discussed.

This work is organized as follows. In Sec. \ref{SummaryTheory} a brief summary of the employed theory for the solution of the electron Boltzmann equation inside the plasma, as well as for the calculation of the SSK is given; the used notation is introduced. Details on the expansion and manipulation of the Boltzmann equation can be found in Appendix \ref{DetLegendre}. Section \ref{BC} presents the construction of the new boundary condition utilizing the SSK in the framework of the expansion introduced earlier. The description of the concrete numerical implementation is delegated to Appendix \ref{NumImpl} and the necessary extra- and interpolation of the SSK, which connects the calculated to experimentally measured values \cite{10.1063/1.4984761} for the secondary electron emission yields (EEYs), can be found in Appendix \ref{Extrapolation}. Numerical results for the new boundary condition implemented into a simulation of an anomalous glow discharge in contact with a silicon anode are presented for different gases in Sec. \ref{Results}. These results are compared to those using simpler boundary conditions. Additionally, an effective reflection coefficient is defined, which gives a quantitative comparison to simpler models and opens a path for usage of the SSK in a simplified way, but also points out the need for an inelastic boundary condition, if good quantitative results are required or secondary electron effects play a major role in the discharge. We conclude in Sec. \ref{Conclusion}.

\section{Theoretical Background}
\label{SummaryTheory}
\subsection{Legendre polynomial expansion of the Boltzmann transport equation}
\label{LegendereExpansion}
\begin{figure}
\includegraphics[width=\linewidth]{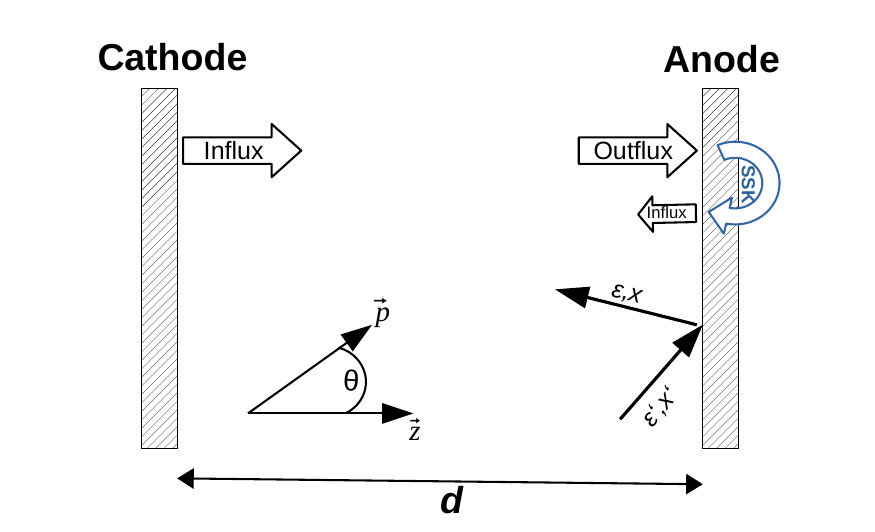}
\caption{Sketch of the geometry of the modeled discharge. The discharge has plane electrodes, a rotational symmetry around the $z$-axis and is axially inhomogeneous. At the cathode the influx of electrons with a given energy profile is used to construct a boundary condition, whereas at the anode the influx into the system is related to the outflux via the SSK.\label{Geometry}}
\end{figure}

The kinetics of the electrons inside a plasma can be described by the Boltzmann transport equation (BTE). The electron momentum distribution function (EMDF) $\tilde{f}(\vec{r},\vec{p})$ of a plasma in a steady state is governed by \cite{Becker2010}
\begin{align}
&\left( \dfrac{\vec{p}}{m_\mathrm{e}} \cdot \nabla_{\vec{r}} + \vec{F} \cdot \nabla_{\vec{p}} \right) \tilde{f}(\vec{r},\vec{p})\nonumber\\
=& \sum\limits_{a} \left( \mathbb{C}_a^{\mathrm{el}}[\tilde{f}] + \sum\limits_{i} \mathbb{C}_{a,i}^{\mathrm{in}}[\tilde{f}] \right),
\label{BTE}
\end{align}
where $m_\mathrm{e}$ is the electron mass, $\vec{F}$ the electric force acting on an electron and $a$ the index of the heavy species. The collision integrals $\mathbb{C}_a^{\mathrm{el}}[\tilde{f}]$ and $\mathbb{C}_{a,i}^{\mathrm{in}}[\tilde{f}]$ describe elastic scattering and the $i$th inelastic scattering process with heavy species $a$, respectively. Equation \eqref{BTE} is now being applied to a dc-discharge with plane parallel electrodes and a rotational symmetry around the $z$-axis (see Fig. \ref{Geometry}). Thus, the electric field $\vec{E}$, the resulting electric force $\vec{F}$ and the plasma inhomogeneity are along the $z$-axis. The dependency of the EMDF can therefore be reduced to $\tilde{f}(z,p,x)$, where $x=\cos\theta$ is the direction cosine between $\vec{p}$ and the $z$-direction.

The EMDF is being expanded in terms of the Legendre polynomials $P_n(x)$:
\begin{equation}
\tilde{f}(z,p,x) = \sum\limits_{n=0}^\infty \tilde{f}_n(z,p)P_n(x).
\end{equation}
Legendre polynomial expansion techniques to solve BTEs have been studied extensively \cite{winklerwilhelmhess,winklerpetrov,wilhelmundwinkler,allis,Grubert2007,Grubert_2014,grubert,Petrov_1997,Boyle_2017}. The details of the substitution of the polynomial expansion into (\ref{BTE}) can be found in Refs. \onlinecite{Becker2010,Grubert_2014,allis}. One proceeds by transforming to the kinetic energy $U=p^2/(2m_\mathrm{e})$ and defining the new expansion coefficients as
\begin{equation}
f_n(z,U) = 2\pi(2m_\mathrm{e})^\frac{3}{2} \tilde{f_n}(z,p(U)).
\end{equation}
To improve the numerical stability, another transformation to the total energy $\varepsilon = U + W(z)$, where $W(z)$ is the potential energy of the electron in the electric field, is performed \cite{Grubert2007}. The collision integrals are partly transformed to a derivative in energy by performing a series expansion in terms of ${m_\mathrm{e}}/{M_a}$ with $M_a$ being the mass of the heavy species \cite{grubert}. This leads to a hierarchy of differential equations \cite{Petrov_1997}
\begin{align}
&\dfrac{n}{2n-1}\left(U\dfrac{\partial}{\partial z} - \dfrac{n-1}{2}F_z\right)f_{n-1}(z,\varepsilon)\nonumber\\ 
+&\dfrac{n+1}{2n+3}\left(U\dfrac{\partial}{\partial z} + \dfrac{n+2}{2}F_z\right)f_{n+1}(z,\varepsilon) \label{boltzmann_legendre_gesamtenergie}\\
	=& 2\sum_{a} N_a\dfrac{m_\mathrm{e}}{M_a}\dfrac{\partial}{\partial \varepsilon}\left(U^2Q_a^\mathrm{d}f_n(z,\varepsilon)\right)\delta_{0n} + C{~^\mathrm{r}}[\vec{f}], \nonumber
\end{align}
where $N_a$ is the density of the heavy particle of kind $a$, $Q_a^\mathrm{d}$ is the momentum transfer cross section for elastic scattering with particles of kind $a$ and  $C^{~\mathrm{r}}[\vec{f}]$ is the remainder of the collision integral, which is not transformed into a derivative. The vector $\vec{f}$ consists of all expansion coefficients $f_n$ up to the expansion order. The complete expressions are given in Appendix \ref{DetLegendre}. For an expansion up to order $l$ these coupled equations have to be solved for $0\leq n \leq l-1$. The coefficients $f_{-1}$ and $f_l$ are set to zero. 

The system is solved on a $(z,\varepsilon)$-grid. The spatial derivatives are taken in-between grid points. For the odd coefficients, determining the moments of the flux, a forward derivative is employed. For the even coefficients, determining the moments of the density, a backward derivative is employed. This leads to a Crank-Nicolson-like scheme \cite{CrankNicolson}. Therefore the odd coefficients need a boundary condition at the left spatial margin ($z=0$) and the even coefficients at the right ($z=d$). The system gets solved ''top-down'' in energy-space, starting at highest energy, where all coefficients are negligibly small to be set to zero, and ending at the lowest simulated energy.

\subsection{Calculation of the Surface Scattering Kernel}
The surface scattering kernel $R(U',x'|U,x)$ is the integration kernel of the integral relation between the EMDF leaving the plasma $f^>(U',x')$ and the EMDF reentering the plasma $f^<(U,x)$. The primed quantities correspond to the electrons leaving the plasma and hitting the wall, the unprimed quantities describe the electrons leaving the wall and reentering the plasma (see Fig. \ref{Geometry}). This fundamental relation can be written as
\begin{equation}
f^<(U,x) = \int\limits_0^\infty \d U' \int\limits_0^1 \d x' ~ R(U',x'|U,x) f^>(U',x').
\label{SSKdef}
\end{equation}
For electrons impinging on semiconductors like silicon, which is used as an exemplary wall material in this work, the calculations can be found in Ref. \onlinecite{PhysRevE.110.035207}.

The computation of the SSK is based on the invariant embedding principle \cite{Dashen1964,GLAZOV2007638} and numerical strategies originally developed for applications in nuclear reactor physics \cite{Shimizu196657}. It utilizes the fact that the physics does not change if an infinitesimally small layer of the same material is added onto an infinite halfspace. Because the mean free path of electrons in a solid is much smaller than the usual thickness of a wall, the wall is well approximated by a halfspace. One can derive an integral equation for the backscattering function by considering the scattering probabilities of all additional paths the electron can take in the added layer of the halfspace. The scattering probabilities considered are for collisions with longitudinal optical phonons, with ion cores and for impact ionization. The ion cores are---accordingly to a randium jellium model---assumed to be randomly distributed in the solid and are represented by phenomenologically screened \cite{Srinivasan1969} pseudo-potentials \cite{PhysRevB.12.4200,PhysRevB.18.4172}. Additionally, image charge effects and reflection at the potential step at the interface were included.

The equation can be solved numerically without further simplifications and yields the SSK for all angles. Because some assumptions in the model do not apply to all energies, the SSK can only be calculated up to an energy of $U\approx30$ eV and has to be extrapolated to higher energies for use as a boundary condition of a plasma (cf. Appendix \ref{Extrapolation}).

\section{Construction of the Boundary Condition}
\label{BC}
The transport equation (\ref{boltzmann_legendre_gesamtenergie}) is a set of $l$ first order differential equations in space. Because of the numerical scheme, we need to define $l/2$ boundary conditions at the anode and $l/2$ boundary conditions at the cathode for each energy step. For both electrodes, the boundary conditions are derived as a generalization of the work of \textsc{Marshak} \cite{Marshak}.

In general, to obtain a physically meaningful and experimentally verifiable boundary condition, one has to find suitable macroscopic quantities characterizing the physics at the boundary and link them to the microscopic expansion coefficients $f_n$. The macroscopic property used here is the $k$th moment of the EMDF in $z$-direction, given by \cite{winklerwilhelmschueller}
\begin{align}
\mathbb{G}_k(z) &= \int\limits_0^\infty  \d p~ p^2  \int\limits_{-1}^1 \d x ~  v_z^k(p) \tilde{f}(z,p,x) \label{kMoment}\\
&= \dfrac{1}{2} \left(\dfrac{2}{m_\mathrm{e}}\right)^{\frac{k}{2}} \int\limits_0^\infty \d U ~ U^{\frac{k+1}{2}} \int\limits_{-1}^1 \d x ~ x^k f(z,U,x). \nonumber
\end{align}
Because the macroscopic quantities do not solely depend on the even or the odd expansion coefficients, it is not possible to set the even/odd coefficients to a fixed value independently of the odd/even coefficients. Instead, one has to look at the coefficients of both parities simultaneously and choose the values of the coefficients acting as boundary condition depending on the coefficients that are calculated with the Boltzmann equation inside the system.

In the following, the boundary condition at the anode, which describes the backscattering of electrons governed by the SSK, and the boundary condition at the cathode, describing a constant influx of electrons into the system, are derived. Both derivations are presented in terms of the kinetic energy $U$, because it is the energy that appears naturally in all quantities in the calculation. The derived relations can be easily transferred to the total energy, used in the numerical solution, by setting $U =\varepsilon-W(z=0)$ at the cathode and $U =\varepsilon-W(z=d)$ at the anode.

\subsection{Backscattering at the anode}
\begin{figure}
\includegraphics[width=\linewidth]{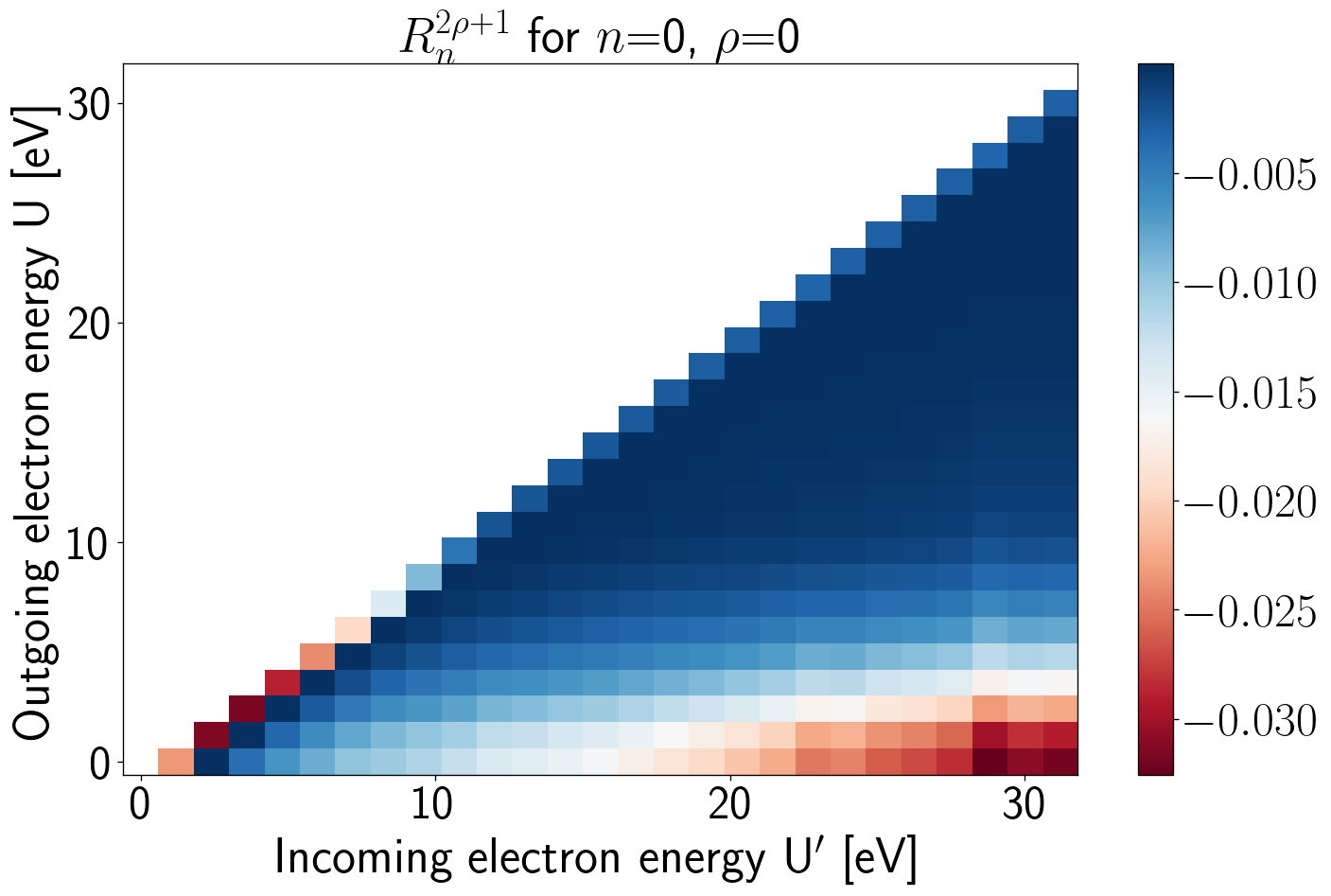}
\caption{The derivate $R_n^{2\rho+1}$ of the SSK is shown for the calculated energy range and one index combination. For other indices, the same energy-dependencies occur, but the sign and the magnitude may change severely.\label{fig:SSK}}
\end{figure}
The inelasticity of the SSK has to be taken into account when deriving the boundary condition with it. Therefore, one must not look at the complete moment as given in (\ref{kMoment}), but at the moment at energy $U$ \footnote{This is in contrast to the approach in Ref. \onlinecite{Becker2010}.
If one uses the approach presented here for the previous boundary condition with a constant reflection coefficient, the same expression is derived in the end, but without the need for a further approximation (Becker et al. had to assume at one point that equality of integrals leads to equality of integrands). We therefore believe that this calculation is correct and ''more rigorous''.}. The starting point are the odd (flux-like) moments of (\ref{kMoment}) at some energy $U$, but only considering electrons with negative velocity, that is $x<0$. If all constant prefactors are dropped, this yields
\begin{equation}
\mathbb{J}^{2\rho+1}_\leftarrow (d,U) = \int\limits_{-1}^0\d x ~x^{2\rho+1} f(d,U,x)
\label{leftMomentAnode}
\end{equation}
with $\rho= 0,\ldots,\frac{l}{2}-1$. The expansion of $f$ in Legendre polynomials leads to
\begin{equation}
\mathbb{J}^{2\rho+1}_\leftarrow (d,U) = \sum\limits_n N_n^{2\rho+1}f_n(d,U),
\label{leftMomentAnodeLeg}
\end{equation}
where
\begin{align}
N_n^{2\rho+1} &= \int\limits_{-1}^0 \d x ~ x^{2\rho+1}P_n(x) \\
&= (-1)^{n+1} \dfrac{(2\rho+1)!}{(2\rho -n + 1)!!(2\rho+n+2)!!} \nonumber
\end{align}
was defined with the double factorials $(\dots)!!$, which are discussed---including the evaluation of negative values---in Ref. \onlinecite{Becker2010}.

These moments can also be calculated using (\ref{SSKdef}):

\begin{align}
&\mathbb{J}^{2\rho+1}_\leftarrow (d,U)\\ &= \int\limits_{-1}^0\d x ~x^{2\rho+1} \int\limits_0^\infty \d U' \int\limits _0^1 \d x' R(U',x'|U,x) f(d,U',x'). \nonumber
\end{align}
Note, that the outgoing direction cosine $x$ was defined with the opposite sign in Ref. \onlinecite{PhysRevE.110.035207}. Now, $f$ is expanded again, yielding
\begin{equation}
\mathbb{J}^{2\rho+1}_\leftarrow (d,U) = \sum\limits_n \int\limits_0^\infty \d U' ~ R_n^{2\rho+1}(U'|U) f_n(d,U')
\label{leftMomentSSKLeg}
\end{equation}
with
\begin{equation}
R_n^{2\rho+1}(U'|U) = \int\limits_{-1}^0\d x ~x^{2\rho+1} \int\limits _0^1 \d x' R(U',x'|U,x) P_n(x').
\end{equation}
By equating (\ref{leftMomentAnodeLeg}) and (\ref{leftMomentSSKLeg}) one finds
\begin{equation}
 \sum\limits_n N_n^{2\rho+1}f_n(d,U) = \sum\limits_n \int\limits_0^\infty \d U' ~ R_n^{2\rho+1}(U'|U) f_n(d,U').
\label{anodeBCcont}
\end{equation}

The quantities $R_n^{2\rho+1}(U'|U)$, which are derived from the SSK, are the actually required data for the implementation of the boundary condition. Example data for $n=0$ and $\rho=0$ are displayed in Fig. \ref{fig:SSK}. For this index combination, the resulting values are all negative as an effect of the sign of the influx. Moreover, it can be seen that elastic backscattering---corresponding to the values on the diagonal---still plays a role and it is stronger than backscattering to medium energies, but much smaller than the backscattering to low energies. This emission at low energies is mainly due to secondary electrons which were excited form the valence band into the conduction band by impact ionization. It becomes important at energies of $\sim 5$ eV and the range to which the secondaries are excited grows slower than linear with the incoming energy of the primary electron.

To make the numerical implementation of (\ref{anodeBCcont}) feasible, the projection of the SSK $R_n^{2\rho+1}(U'|U)$ needs to be split into an elastic and an inelastic part:
\begin{equation}
R_n^{2\rho+1}(U'|U) = r_n^{2\rho+1}(U)\delta(U'-U)+\Delta R_n^{2\rho+1}(U'|U).
\end{equation}
The elastic scattering coefficient $r_n^{2\rho+1}(U)$ is the equivalent to the elastic reflection coefficient of the previous boundary condition \cite{D_Loffhagen_2001,D_Loffhagen_2002,Becker2010}, but it has a $\rho$-dependency due to the angle-dependency of the SSK. The boundary condition can finally be written as
\begin{equation}
\sum\limits_n \left( N_n^{2\rho+1} - r_n^{2\rho+1}(U) \right) f_n(d,U) = S^{2\rho+1}(U),
\label{finalAnodeBC}
\end{equation}
with
\begin{equation}
S^{2\rho+1}(U) = \sum\limits_n \int\limits_0^\infty \d U' ~ \Delta R_n^{2\rho+1}(U'|U) f_n(d,U').
\end{equation}
This quantity describes the inelastically backscattered and emitted secondary electrons and has no analogue in the previous boundary condition. Note, that $S^{2\rho+1}(U)$ is independent of $n$ and therefore the same for all $f_n$. The numerical implementation of (\ref{finalAnodeBC}) is described in Appendix \ref{NumImpl}.

\subsection{Influx at the cathode}
At the cathode, a constant influx of electrons is assumed. This influx is described by a source function $f_\mathrm{S}(U,x)$. The source function assumed in this work is isotropic in the positive halfspace and has a Gaussian-like energy profile, given by
\begin{equation}
f_\mathrm{S}(U,x) = \hat{f_\mathrm{S}} \cdot U \exp \left(-\left(\dfrac{U-U_\mathrm{c}}{U_\mathrm{w}}\right)^2\right)
\label{SourceFunc}
\end{equation}
with $U_\mathrm{c}=5$ eV, $U_\mathrm{w}=1.5$ eV and $\hat{f_\mathrm{S}}$ being a normalization factor. The physical condition at the boundary is
\begin{equation}
f(0,U,x) = f_\mathrm{S}(U,x),
\end{equation}
which is used with the moments (\ref{kMoment}). Analogously to (\ref{leftMomentAnode}), the moments pointing away from the cathode are defined as
\begin{equation}
\mathbb{J}^{2\rho+1}_\rightarrow (0,U) = \int\limits_{0}^1\d x ~x^{2\rho+1} f(0,U,x).
\end{equation}
By plugging in the Legendre polynomial expansion for $f(0,U,x)$ one finds
\begin{equation}
\mathbb{J}^{2\rho+1}_\rightarrow (0,U) = \sum\limits_n (-1)^{n+1} N_n^{2\rho+1} f_n(0,U)
\end{equation}
whereas using $f_\mathrm{S}(U,x)$ leads to
\begin{equation}
\mathbb{J}^{2\rho+1}_\rightarrow (0,U) = \int\limits_0^1\d x ~ x^{2\rho+1} f_\mathrm{S}(U,x) = \mathbb{J}_\mathrm{S}^{2\rho+1}.
\end{equation}
The influx boundary condition at the cathode then becomes \cite{Becker2010}
\begin{equation}
\sum\limits_n (-1)^{n+1} N_n^{2\rho+1} f_n(0,U) = \mathbb{J}_\mathrm{S}^{2\rho+1}.
\end{equation}
These expressions can be evaluated for all $\rho=0,\dots, l/2$ or only for $\rho=0$ as it is done here.

\section{Numerical results}
\label{Results}

\subsection{Effects for an Ar discharge}
The effects of the altered boundary condition at the anode are being investigated for an anomalous Ar glow discharge. Simulations were run for the new boundary condition based on the SSK and the old boundary condition with a constant, elastic reflection coefficient $r$. The latter was performed with a value of $r=0.36$, as used in previous publications \cite{Petrov_1997,Becker2010,grubert} and with $r=0$, realizing a perfect absorber. The discharge parameters \cite{D_Loffhagen_2002} can be found in Table \ref{table_parameters}. The BTE is not solved self-consistently with the Poisson equation, but depending on a parameterized potential $V(z)$, which was calculated in Ref. \onlinecite{Grubert_2014}. The densities of the heavy species were calculated from the ideal gas law. For all simulations expansion order $l=8$ was used.

\begin{table}
	\caption{The discharge parameters used in the different simulations. The parameters were chosen such that comparable anomalous glow discharges are experimentally achievable.}
	\begin{tabular}{c|c|c|c}
		& set 1 & set 2 & set 3 \\
		\hline\hline
		gas&  Ar &  He &  O$_2$ \\
		pressure $p$ [Pa]& 2.5 & 10 & 5 \\
		system length $d$ [cm]& 10 & 15 & 10 \\
		current density [$\mathrm{\mu}$A/cm$^2$]& 0.82 & 0.37 & 2.98 \\
		cathode fall voltage [V]& 300 & 300 & 300 \\
		cathode fall thickness [cm]& 7.74 & 13.55 & 6.94 \\
		ion density (ground state) [$10^{15}$ cm$^{-3}$]& 0.6 & 2.41 & 1.21 \\
		ion density (excited state) [$10^{10}$ cm$^{-3}$]& 0.6 & 2.41 & 1.21 \\
		\hline\hline
	\end{tabular}
	\label{table_parameters}
\end{table}

\begin{figure}
\centering
\includegraphics[width=\linewidth]{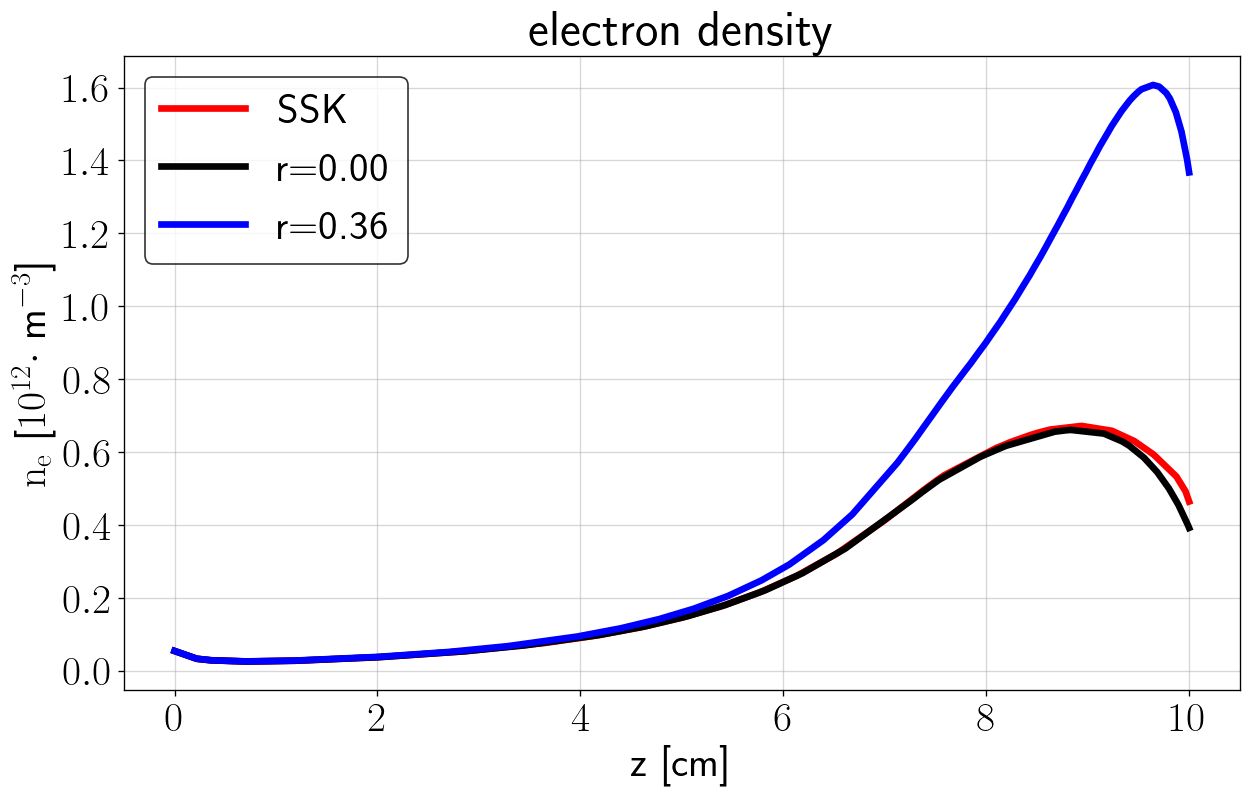}
\caption{The total electron density is plotted over the position $z$ for parameter set 1 (Ar).}
\label{fig:TotalDens}
\end{figure}
\begin{figure}
\centering
\includegraphics[width=\linewidth]{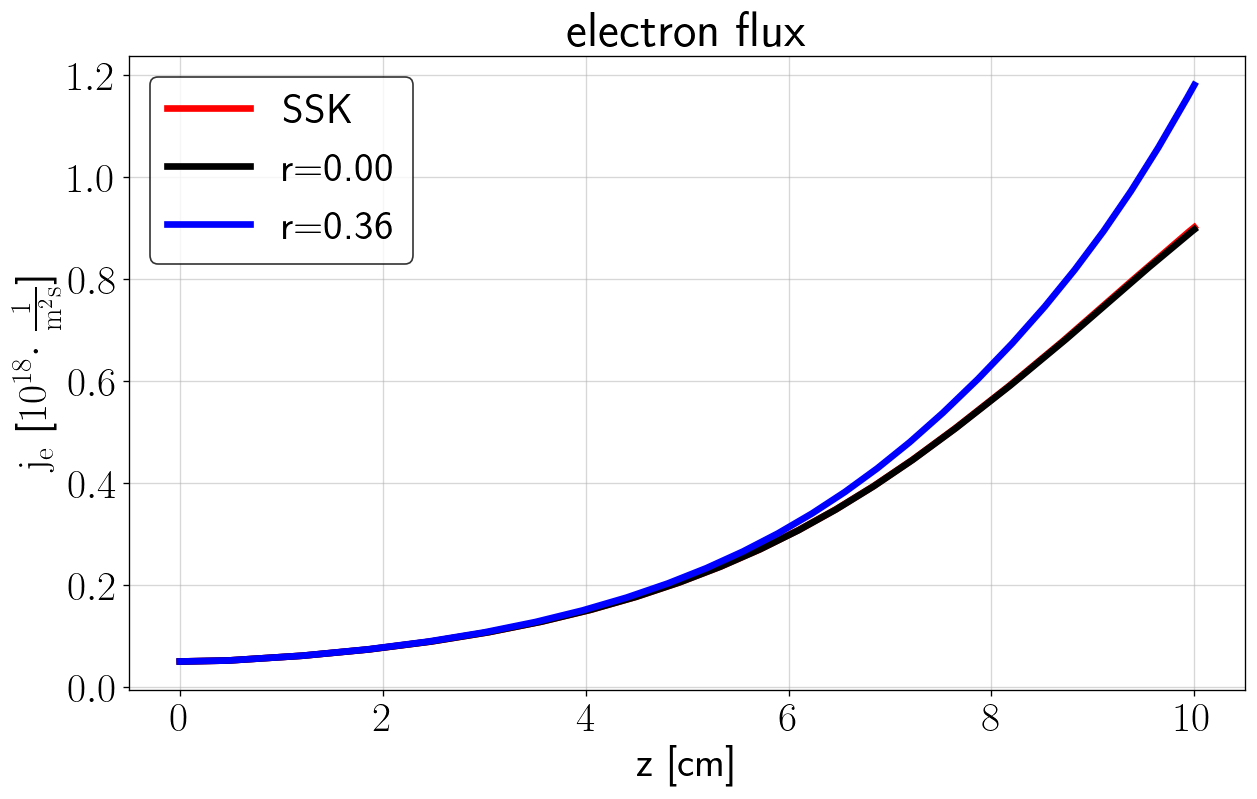}
\caption{The total electron flux in the z-direction is plotted over the position $z$ for parameter set 1 (Ar).}
\label{fig:TotalFlux}
\end{figure}

At first, the total electron densities and fluxes are compared for the different boundary conditions. The electron density $n_e(z)$ (Fig. \ref{fig:TotalDens}) shows that the increased reflection leads to an increased density, not only in the sheath, but also in the bulk. Interestingly, for Si the results for the SSK-boundary condition are closer to the perfect absorber than to those of the previously used boundary condition. When looking at the flux $j_z(z)$ in the $z$-direction (Fig. \ref{fig:TotalFlux}) one sees the same, but the relative differences are even smaller. These deviations can be explained by looking at energy-resolved quantities.

\begin{figure*}
\centering
\includegraphics[width=\textwidth]{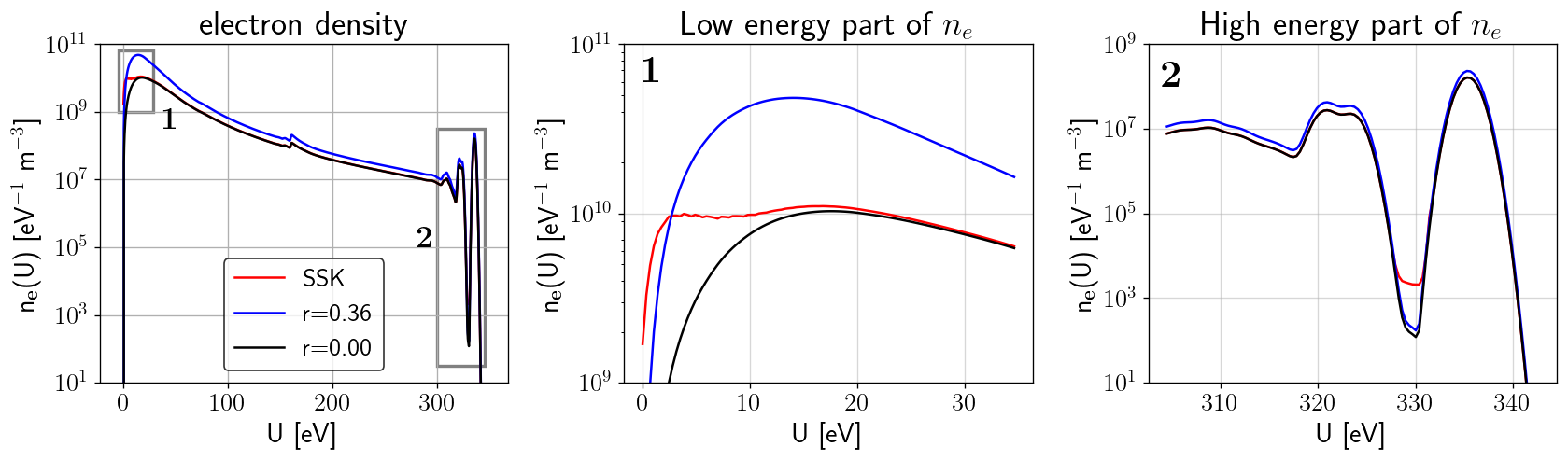}
\caption{The electron density at the anode ($z=d=10$ cm) is shown on a logarithmic scale. The middle and the right plot are zoomed in parts of the first one.}
\label{fig:densAtAnode}
\end{figure*} 
\begin{figure*}
\centering
\includegraphics[width=\textwidth]{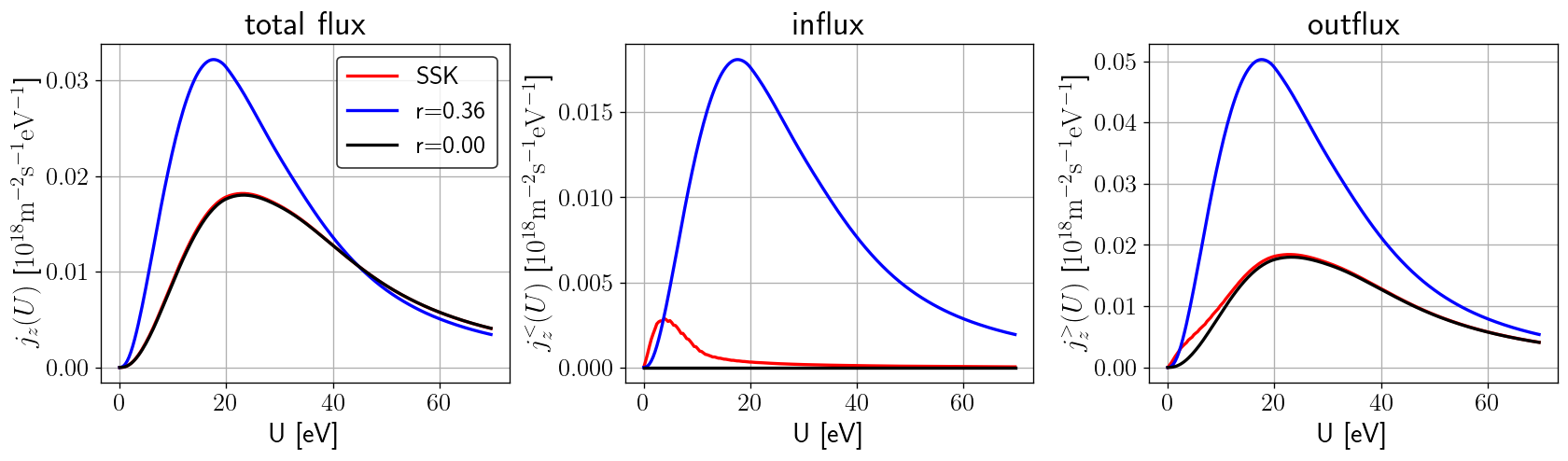}
\caption{The electron flux at the anode is shown. The energy range is the one of the low energy part in Fig. \ref{fig:densAtAnode}. The influx and outflux are defined from the plasma's point of view, cf. Fig \ref{Geometry}, but neglecting the negative sign of the influx.}
\label{fig:fluxAtAnode}
\end{figure*}

By looking at the EMDF directly at the anode ($z=d$) the energetic effects of the new boundary condition become clear. First, the electron density $n_e(d,U)$ at the anode (Fig. \ref{fig:densAtAnode}) is investigated. The electron density covers several orders of magnitude and is much higher at low energies than at high energies, which is dominantly an effect of the thermalization in the plasma, but also the inelasticity of the boundary condition plays a role. The two peaks at high energies stem from the Gaussian at the cathode and represent electrons that travel to the anode without any, respectively, exactly one inelastic scattering event(s). The kink at $\sim 165$ eV exists due to the modeling of the ionization, where it is assumed that the energy is distributed equally to both electrons. The steep rise towards lower energies is an effect of the plasma sheath.

The effects of the new boundary condition are best seen in the very high and very low energy region. At low energies the difference to the perfect absorber becomes clear. The increased electron emission at low energies (due to electrons impinging at higher energies) results in the density being many times higher than the densities of the perfect absorber in the region of about 2 eV. This strong increase in relative numbers is important, because the densities are very high in absolute numbers at these energies. The densities with the new boundary condition even get higher than the densities using $r=0.36$. At high energies the inelasticity of the new boundary condition leads to a smoothening of the EMDF. The region in-between the two peaks, where through processes in the plasma almost no electrons can appear, gets filled up by electrons, which hit the anode in the region of the higher peak and are scattered inside the wall toward lower energies. 

These effects explain the behaviour of the flux at the anode at low energies (Fig. \ref{fig:fluxAtAnode}). When only the total flux is looked at, the perfect absorber and the new boundary condition do not differ significantly in contrast to the density. However, for the influx back to the system, differences can be seen. The perfect absorber does not produce an influx, while the SSK-boundary condition produces a peak shifted to lower energies when compared to the elastic boundary condition with $r=0.36$. The increased influx also leads to an increased outflux, because the reentering electrons can scatter in the plasma and reach the anode again. These two increased fluxes cancel each other in the total flux. The increased density, in contrast, is caused by electrons moving towards and away from the anode almost equally.

\subsection{Comparison between gases}

\begin{figure*}

\includegraphics[width=\textwidth]{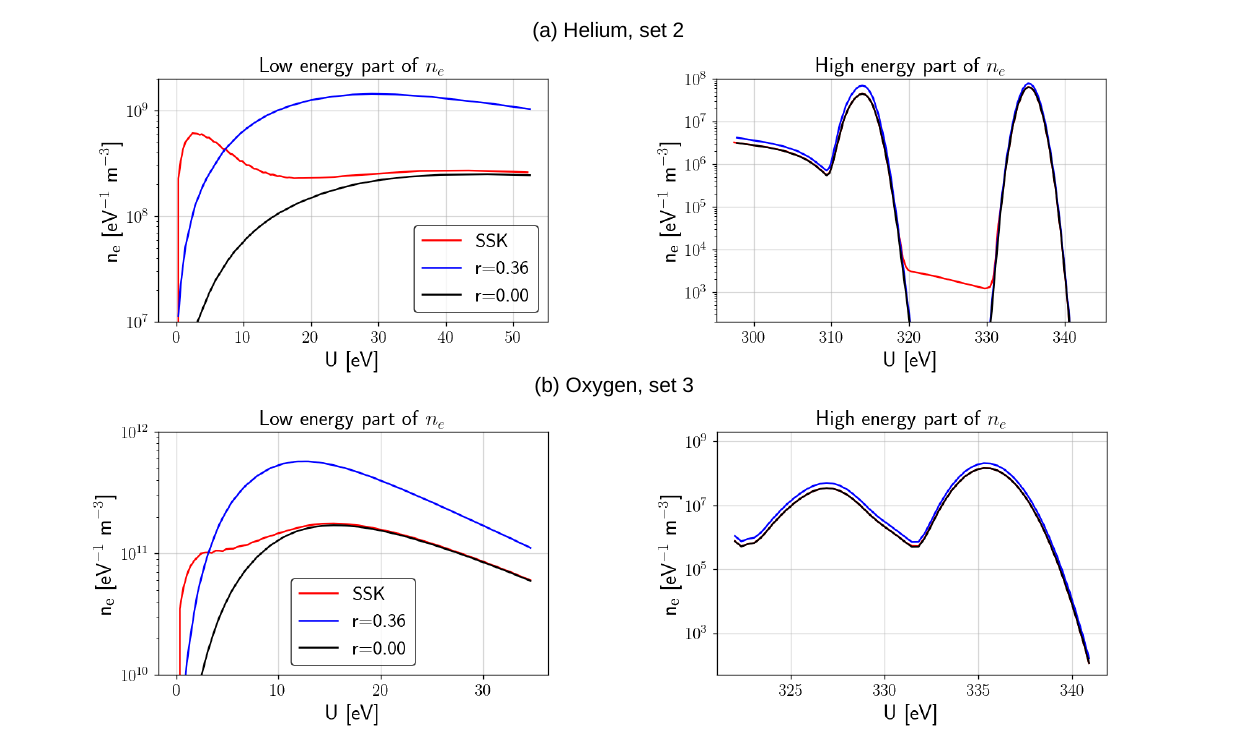}

\caption{\label{fig:densOxyHe}The density profiles for the parameter sets 2 and 3 at the anode ($z=d$). The plotted energy ranges are chosen such that the same physical effects can be seen as in the plots of Fig. \ref{fig:densAtAnode}.}

\end{figure*}

The results of the argon discharge from the previous section are being compared with discharges using helium (He) and oxygen ($\mathrm{O_2}$) as a background gas. As one can see from Table \ref{table_parameters}, the parameters of these three plasmas are quite different, because different gases need different conditions to ignite a discharge. Hence, only data directly at the anode will be compared qualitatively and the energy regions that are studied in detail are shifted.

In Fig. \ref{fig:densOxyHe} the density profiles can be found. A complete discussion of the reaction dynamics in these discharges is given in Ref. \cite{grubert}. Here, only the changes due to the boundary condition are discussed. In the low energy range a similar picture compared to the Ar discharge appears: The increase in density with the new boundary condition is seen for all three gases, but for He the highest relative increase can be seen. This is due to the flatter rise of the density in this energy regime. Therefore, the elastic backscattering becomes less important and the inelastic effects of the SSK-boundary condition are better visible. In the high energy range the differences between the gases are more severe: While for Ar the smoothening of the density was effective for a small energy range ($\Delta U \approx 3$ eV) and for non-vanishing densities, it is effective for He for a much larger energy range ($\Delta U \approx 10$ eV) and increases an elsewise vanishing density, whereas for O$_2$ almost no effect is visible. This can be explained with the minimal excitation energies of the molecules. The O$_2$ molecule has vibrational modes with very low excitation thresholds, while Ar atoms do not have vibrational modes and therefore much higher energy thresholds. He atoms have even higher ones. Hence, there are no 'forbidden' zones for O$_2$, which could be filled via inelastic wall processes and for He those 'forbidden' zones are much wider and even wider than the peak of the source function (\ref{SourceFunc}) at the cathode. The improved boundary condition is therefore most relevant for gases with high excitation thresholds.

\subsection{Effective reflection coefficient}
\begin{figure}
\centering
\includegraphics[width=\linewidth]{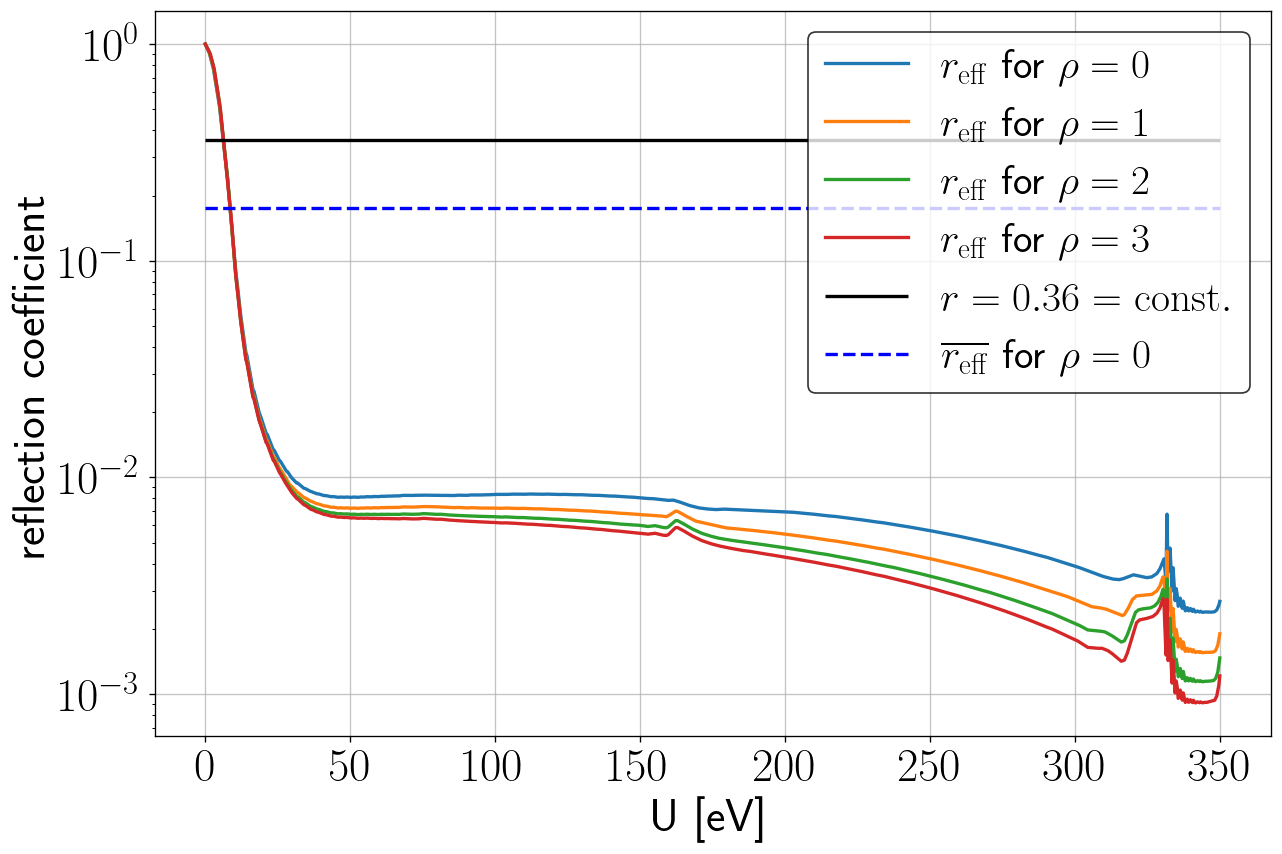}
\caption{The effective reflection coefficient $r_\mathrm{eff}$ is plotted for different orders $2\rho + 1$ over the energy and compared to the constant reflection coefficient $r=0.36$. The results were obtained with a wider influx profile ($U_w = 4$ eV) than usual to avoid extreme peaks when the density gradients become too high, as explained in the main text.}
\label{fig:effR}
\end{figure}

In the following, an effective reflection coefficient is introduced. Mainly, to have a simple metric to measure the strength of the backscattering and to compare it quantitatively with the constant-r boundary condition. Especially, the weighted average of the effective reflection coefficient is closely related to the constant $r$. This is also the most direct way to get a simplified boundary condition if one is interested in using the data from the SSK, but either does not want to or cannot implement the inelastic processes at the boundary.

This approximation would mean using in- and outflux profiles from a given simulation, calculating the effective reflection coefficient from it and using it for a more complicated simulation of a discharge with similar electron distributions and, therefore, comparable electron producing reactions. Thereby, inelastic quantities are artificially projected onto the elastic coefficients. The effective reflection coefficient is hence not an advised approximation, but rather illustrates well how strong the inelasticity is and that neither an elastic approach nor a perfect absorber is a good approximation.

The calculation of the effective reflection coefficient is motivated by the mathematical description of the boundary condition with constant reflection coefficient $r$. It is given by
\begin{equation}
\mathbb{J}^{2\rho+1}_\leftarrow (d,U) = -r\cdot \mathbb{J}^{2\rho+1}_\rightarrow (d,U),
\end{equation}
where the minus sign represents the change of direction. This immediately gives rise to the definition of the effective reflection coefficient
\begin{equation}
    r_\mathrm{eff}^{2\rho+1}(U)= -\dfrac{\mathbb{J}^{2\rho+1}_\leftarrow (d,U)}{\mathbb{J}^{2\rho+1}_\rightarrow (d,U)}.
\end{equation}
It is not only energy-dependent, but also depends on the order of the moment, since the SSK is also angle-dependent. The coefficient $r_\mathrm{eff}^1$ is physically the most meaningful, because it relates the outflux to the influx.

The effective reflection coefficient for the Ar discharge (set 1) can be found in Fig. \ref{fig:effR} up to order 7. One sees that $r_\mathrm{eff}(U)$ gets close to unity at very low energies, while it is only a few percent at higher energies. This is due to the inelasticity and the influx into the plasma being mostly at low energies. At $\sim 165$ eV and $\sim 330$ eV small peaks are located. They are an effect of comparing only moments at the same energy, while the underlying physics is more complex. These peaks become larger, when the slopes of the profiles of density, flux, etc. become steeper. This underlines that this approximation is to be used with care.

Also plotted are the constant reflection coefficient $r=0.36$ and the weighted average $\langle r_\mathrm{eff}^1\rangle\approx 0.175$. This means that for the case of Si approximately 17.5 \% of the electron flux hitting the anode is being backscattered, which should post a better approximation as a constant reflection coefficient $r=0$ or $r=0.36$. The differences between the different order moments are much smaller than the differences between different energies. Accordingly, approximations neglecting the angle-dependence of the electron-wall-interaction can be justified better than approximations neglecting the inelasticity.

\subsection{The role of pressure for the importance of the boundary condition}
\begin{figure}
\includegraphics[width=\linewidth]{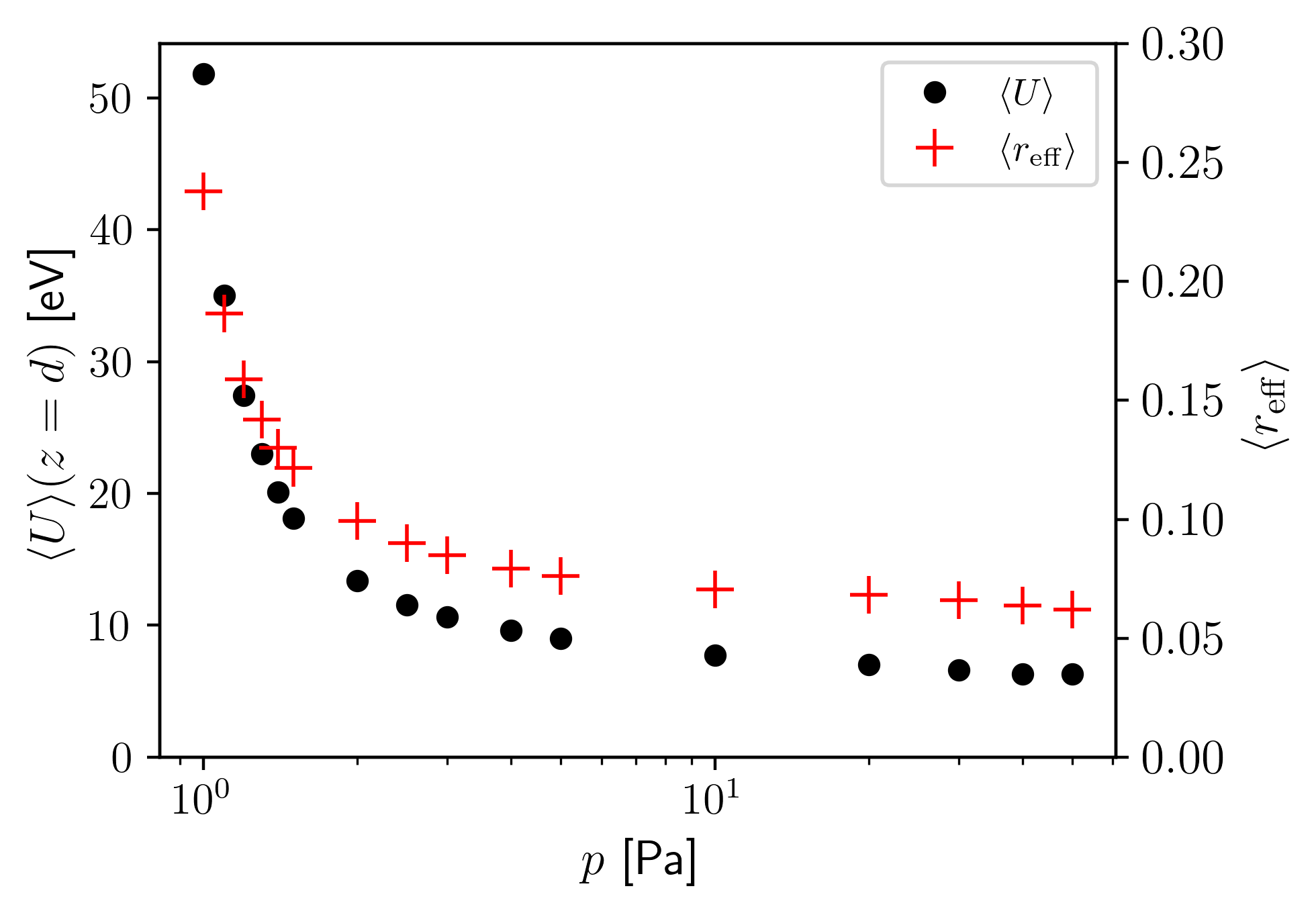}
\caption{\label{UmeanReff}The average kinetic energy of the electron density directly at the anode and the weighted average of the effective reflection coefficient are shown for different pressures. The results were obtained for an Ar discharge with size $d=22$ cm. All discharge parameters except the pressure were kept constant and are comparable to those of set 1. It becomes clear that the mean energy and the mean effective reflection coefficient are highly correlated.}
\end{figure}

As it was seen before, the total electron density at the anode is closer to the perfect absorber than to the previously used boundary condition with $r=0.36$. It is not clear yet, under which conditions the perfect absorber can be a reasonable approximation and under which conditions a correct boundary condition is necessary to get good results. A correct description of the plasma-wall-interaction is important, if the electron emission is strong, which is the case if the EEYs are high. In the energy regime here, the EEYs increase with increasing energy. Therefore, the SSK-boundary condition is needed in regimes where the average energy of the electrons at the boundary is relatively high.

To increase the electron energy at the anode, one can either increase the energy of the injected electrons, increase the voltage or decrease the pressure. The first two options do not lead to a relevant increase of the electron energy at the anode, because the electron energy is mostly increased in the region of the cathode fall only, but the energy gain there almost gets compensated completely by higher energy losses in scattering events, whereas decreasing the pressure has a big influence on the electron energy, because it decreases the number of scattering events in the plasma and with it the average energy loss of an electron traveling from the cathode to the anode.

Therefore, discharges at different pressures are compared. As a metric for the importance of the SSK-boundary condition the weighted average of the effective reflection coefficient is used. As one can see in Fig. \ref{UmeanReff}, a decrease in pressure leads to an increase of both $\langle U \rangle (z=d)$ and $\langle r_\mathrm{eff}^1 \rangle$. Also, $\langle U \rangle (z=d)$ and $\langle r_\mathrm{eff}^1 \rangle$ are highly correlated, showing that the electron energy at the boundary is the main factor to determine whether a perfect absorber approximation is useful. At the example of Si, one can say that for low average energies (below 10 eV) the perfect absorber can be a valid approximation for the boundary condition, if one is mainly interested in macroscopic, energy averaged quantities like the total density or flux. For higher energies that correspond for this special discharge to pressures below 2 Pa, even the macroscopic quantities differ significantly from the ones obtained by a perfect absorber approximation and  energy-dependent and inelastic boundary conditions become necessary. For other anode materials and other types of discharges, the concrete values may change drastically.

\section{Conclusion}
\label{Conclusion}
A boundary condition for the electron kinetics of a plasma was derived which is based on the microphysical description of the electron-wall-interaction. The numerical implementation for a multi-term BTE solver was shown, which has proven to be stable and requiring a negligible additional numerical effort.

The numerical results for an anomalous glow discharge verified the need for a boundary condition that accounts for the inelastic processes inside the wall, whereas resolving angular dependencies is secondary for Si and this type of discharge which has highly isotropic EMDFs throughout the system. The changes in the EMDF are obviously most severe near the anode, but also affect the bulk. Hence, the improved boundary condition has the largest impact for discharges, where reactions near the surface play an important role and also for plasmas, where the excitation energies in the plasma are much higher than the energy losses of the electron in the solid, because here the boundary condition leads to a smoothening of the EMDF.

It was shown that an elastic, constant reflection coefficient $r$ is not a good approximation and that any elastic approach cannot describe the physics of the plasma-wall-interaction sufficiently. The common cases of $r=0$ and $r=0.36$ were not reproducible for Si, although macroscopic, energy averaged densities and fluxes are close to the ones obtained by the perfect absorber model. Energy resolved quantities, however, differ significantly. 

The calculation of the SSK, on which the boundary condition is based, is perhaps more demanding than the modeling by phenomenological formulae, but it is not affected by the main problem phenomenological models face: The plasma-wall-interaction needs to be described at least four-dimensionally (in-/outgoing energy and angle). To approximate it phenomenologically requires thus a sufficiently complex function and hence a high number of fitting parameters (for the model in Ref. \onlinecite{PhysRevSTAB.5.124404} over 30). To determine these parameters, a sufficient database is required, but electron emission yields or related data are, to our knowledge, most of the time only available for rather high energies and for normal incidence. Therefore, the sparse data that are available for angle-dependent backscattering needs to be applied to energy ranges that are far away from the measured points, making the obtained function a rough approximation.

The SSK-boundary condition bypasses the lack of experimental data by combining a microphysical wall description at low energies with measured yields at high energies. In the low energy-range, where little to none experimental data are available, the SSK is calculated theoretically. Since the most important energy relaxation process inside the solid is the impact ionization, which leads to emitted electrons at low energies, the angular distribution of the emitted electrons in this energy range is very well approximated by the calculated distribution, even for much higher impact energies (for more details see Appendix \ref{Extrapolation}). Only the energy dependence needs experimental input data for higher energies, but for these energies measured EEYs are available. This leaves only the dependence on the incoming angle at high energies as a loose end.

\appendix
\section{Expansion of the collision integral}
\label{DetLegendre}
All collision integrals are derived under the assumption of a classical two-particle-collision between an electron and a particle of a heavy species. The expressions are further simplified by using $m_\mathrm{e}\ll M_a$. After performing the series expansion in ${m_\mathrm{e}}/{M_a}$ the expression \cite{grubert}
\begin{widetext}
\begin{align}
C^{~\mathrm{r}}\left[\vec{f}\right] = &-\sum_{a}UN_a\left[ Q_a^\mathrm{d}(U)\Theta(n-1)+\sum_{i}Q_{i,a}^\mathrm{in}(U) \right]f_n(z,\varepsilon) \\
	&+\sum_{a}N_a\sum_{i}\beta_{i,a}^2\cdot(\beta_{i,a}U+U_{i,a}^\mathrm{in})\cdot Q_{i,a}^\mathrm{in}(\beta_{i,a}U+U_{i,a}^\mathrm{in})\nonumber
	\times f_n(z,\beta_{i,a}\varepsilon+(1-\beta_{i,a})W+U_{i,a}^\mathrm{in})\delta_{0n} \nonumber
\end{align}
\end{widetext}
for the remainder of the collision integral is obtained. Here, $\Theta$ is the Heaviside function, $\delta_{ij}$ the Kronecker delta, $U_{i,a}^\mathrm{in}$ the threshold energy of the inelastic scattering process $i$. The parameter $\beta_{i,a}$ was introduced to implement the common assumption that the excitation energy of a single particle is transferred in the scattering process equally to both partners. For ionizing collisions $\beta_{i,a}=2$ and for other collisions $\beta_{i,a}=1$. The differential scattering cross sections $\sigma^\mathrm{in/el}(U,\theta)$ enter the collision integrals via
\begin{equation}
Q_{i,a}^\mathrm{in}(U) = \int\mathrm{d}\Omega ~\sigma_{i,a}^\mathrm{in}(U,\theta) P_n(\cos \theta)
\end{equation}
and
\begin{align}
Q_a^\mathrm{d}(U) = &\int\mathrm{d}\Omega~ \sigma_{i,a}^\mathrm{el}(U,\theta) P_n(\cos \theta)\\
 &- \int\mathrm{d}\Omega ~\sigma_{i,a}^\mathrm{el}(U,\theta) P_n(\cos \theta)P_1(\cos\theta). \nonumber
\end{align}
\section{Numerical implementation of the boundary condition}
\label{NumImpl}
The numerical implementation of the BTE solver for the electron kinetics is based on the work of \textsc{Petrov} et al. \cite{Petrov_1997}. It was adapted and advanced by \textsc{Grubert} \cite{grubert}, whose code was used in this work. As this article is concerned with the boundary condition of the electron Boltzmann equation, it is sufficient to understand the structure of the equation system that has to be solved numerically.

As already presented in Sec. \ref{LegendereExpansion}, the BTE is solved ''top-down'' from the highest energy to the lowest, with the total energy acting as a quasi-time. For each energy step $\varepsilon_j$ (the index $j$ will be dropped in this appendix and the index $n$, referring to the Legendre expansion coefficients, will be placed as a left-hand superscript) a linear equation system of the form
\begin{widetext}
\begin{equation}
\begin{bmatrix}
B_0	&	C_0	& 0 & \dots & & &\\
A_1	& B_1	& C_1 & 0	& \dots & &\\
0	& A_2	& B_2	& C_2 & 0 & \dots &\\
&&\ddots&\ddots&\ddots\\
& \dots & 0 & A_{i_\mathrm{max}-1}	& B_{i_\mathrm{max}-1}	& C_{i_\mathrm{max}-1}	\\
&&  \dots & 0 & A_{i_\mathrm{max}}	& B_{i_\mathrm{max}}	
\end{bmatrix}
\cdot
\begin{bmatrix}
\vec{f}_0\\
\vec{f}_1\\
\vec{f}_2\\
\vdots\\
\vec{f}_{i_\mathrm{max}-1}\\
\vec{f}_{i_\mathrm{max}}
\end{bmatrix}
=
\begin{bmatrix}
\vec{\Delta}_0\\
\vec{\Delta}_1\\
\vec{\Delta}_2\\
\vdots\\
\vec{\Delta}_{i_\mathrm{max}-1}\\
\vec{\Delta}_{i_\mathrm{max}}
\end{bmatrix}
\label{MasterEq}
\end{equation}
\end{widetext}
has to be solved for the expansion coefficients $^nf_i$ at position $z_i$, which are joined up in the vector $\vec{f}_{i} = [{^0f}_i,\dots,{^{l-1}f}_i]^T$, whereas the vector $\vec{\Delta}_{i} = [^0\Delta_i,\dots,{^{l-1}\Delta}_i]^T$ consists of all the contributions from energy $\varepsilon_{j+1}$, which has been solved in the previous step. As $\vec{f}_i$ and $\vec{\Delta}_i$ are vectors, $A_i$, $B_i$ and $C_i$ are block matrices of size $l \times l$. 

As only the construction of the boundary condition at the anode deviates from earlier works and the anode is located at the last spatial discretization point $z_{i_\mathrm{max}}$, the only quantities which have to be altered are $B_{i_\mathrm{max}}$ and $\vec{\Delta}_{i_\mathrm{max}}$. The block matrix $A_{i_\mathrm{max}}$ describes the influence of the expansion coefficients at $z_{i_\mathrm{max}-1}$ on the expansion coefficient at $z_{i_\mathrm{max}}$ via the convection term of the BTE and will not be changed by the boundary condition. The matrix $B_{i_\mathrm{max}}$ without the boundary condition is given by
\begin{align}
B_{i_\mathrm{max}} = 
\begin{pmatrix}
0&0&0&0& 0&\dots &&&\\
\Psi & \Gamma & \Phi & 0 & 0 & \dots&&&\\
0& 0&0&0&0&\dots&&&\\
0 & 0 & \Psi & \Gamma & \Phi & \dots&&&\\
&&&&&\ddots&&&\\
&&&&& \dots & 0&0&0&0\\
&&&&& \dots & 0 & 0 & \Psi & \Gamma
\end{pmatrix}
\label{B_imax}
\end{align}
with $\Psi$, $\Gamma$ and $\Phi$ being energy-dependent coefficients from the discretization of the $z$-derivative, but their concrete values are unnecessary for understanding the following. The corresponding inhomogeneity is given by
\begin{equation}
\vec{\Delta}_{i_\mathrm{max}} = 
\begin{pmatrix}
0 & ^1\Delta_{i_\mathrm{max}}&
0&
^3\Delta_{i_\mathrm{max}}&
\dots&
0&
^{l-1}\Delta_{i_\mathrm{max}}
\end{pmatrix}^T.
\label{Delta_imax}
\end{equation}

The rows for even indices are zero, because these rows would be filled with coefficients from the backward derivative for the even indices, if the point was not at the right spatial margin of the system. Without these numbers, the resulting linear equation system is underdetermined. It needs to be filled by the boundary condition.

The boundary condition at energy $\varepsilon_j$, when transformed to total energy, reads \footnote{Using the discrete version of the boundary condition means that $r_n^{2\rho+1}$ does not only include strictly elastic processes, but also processes which lead to an energy loss of less than half the energy spacing, which cannot be resolved with this discretization.}
\begin{equation}
\sum\limits_{n=0}^{l-1} \left( {^n}N^{2\rho+1} - {^n}r^{2\rho+1} (\varepsilon_j) \right) \cdot {^nf}^{i_\mathrm{max}}(\varepsilon_j) = S^{2\rho+1}(\varepsilon_j).
\end{equation}
The inhomogeneity $S^{2\rho+1}(\varepsilon_j)$ only depends on values of $^nf^{i_\mathrm{max}}(\varepsilon_j')$ with $j'>j$. This is because no physical process in the wall leads to a gain of energy for the electrons entering it. Thus, $S^{2\rho+1}(\varepsilon_j)$ can be calculated each step using only values calculated from earlier steps. By defining
\begin{equation}
^nM^{2\rho+1}(\varepsilon_j) = {^nN}^{2\rho+1} - {^n}r^{2\rho+1} (\varepsilon_j),
\end{equation}
dropping the argument and using $\rho = 0,\dots,l/2-1$, one gets $l/2$ equations, which can be translated into another underdetermined linear equation system:
\begin{widetext}
\begin{equation}
\begin{pmatrix}
0&0&0&0& 0&\dots &&&\\
^0M^1 & ^1M^1 & ^2M^1 & ^3M^1 & ^4M^1 & \dots&&&\\
0& 0&0&0&0&\dots&&&\\
^0M^3 & ^1M^3 & ^2M^3 & ^3M^3 & ^4M^3 & \dots&&&\\
&&&&&\ddots&&&\\
&&&&& \dots & 0&0&0&0\\
&&&&& \dots & ^{l-4}M^{l-1} & ^{l-3}M^{l-1} & ^{l-2}M^{l-1} & ^{l-1}M^{l-1} 
\end{pmatrix}
\cdot
\begin{pmatrix}
^0f_{i_\mathrm{max}}\\
^1f_{i_\mathrm{max}}\\
^2f_{i_\mathrm{max}}\\
^3f_{i_\mathrm{max}}\\
\vdots\\
^{l-2}f_{i_\mathrm{max}}\\
^{l-1}f_{i_\mathrm{max}}
\end{pmatrix}
= \begin{pmatrix}
0\\
S^1\\
0\\
S^3\\
\vdots\\
0\\
S^{l-1}
\end{pmatrix}.
\end{equation}
To use this equation in (\ref{MasterEq}), the system needs to fill the zero-rows of (\ref{B_imax}) and (\ref{Delta_imax}) with numbers. Therefore, the zeroth row of the matrix and the inhomogeneity gets interchanged with the first, the second with the third, and so on. This leads to a rather unorthodox notation, but it does not change the solution of the equation system. The interchanged matrix then gets added to (\ref{B_imax}), yielding
\begin{equation}
B_{i_\mathrm{max}} = 
\begin{pmatrix}
^0M^1 & ^1M^1 & ^2M^1 & ^3M^1 & ^4M^1 & \dots&&&\\
\Psi & \Gamma & \Phi & 0 & 0 & \dots&&&\\
^0M^3 & ^1M^3 & ^2M^3 & ^3M^3 & ^4M^3 & \dots&&&\\
0 & 0 & \Psi & \Gamma & \Phi & \dots&&&\\
&&&&&\ddots&&&\\
&&&&& \dots & ^{l-4}M^{l-1} & ^{l-3}M^{l-1} & ^{l-2}M^{l-1} & ^{l-1}M^{l-1} \\
&&&&& \dots & 0 & 0 & \Psi & \Gamma
\end{pmatrix}
\end{equation}

and the inhomogeneity is added to (\ref{Delta_imax}), leading to
\begin{equation}
\vec{\Delta}_{i_\mathrm{max}} = 
\begin{pmatrix}
{S^1}&
^1\Delta_{i_\mathrm{max}}&
{S^3}&
^3\Delta_{i_\mathrm{max}}&
\dots&
{S^{l-1}}&
^{l-1}\Delta_{i_\mathrm{max}}
\end{pmatrix}^T.
\end{equation}
\end{widetext}
These manipulated vectors and matrices allow the solution of the system (\ref{MasterEq}) for each $\epsilon_j$ with respect to the new boundary condition in the same manner as before, that is using a sparse matrix solver to determine the $\vec{f}_{i}$. This means that the additional numerical workload for this new boundary condition compared to the old one is the calculation of $S^{2\rho+1}$ and the reading in of the data for $r_n^{2\rho+1}$ and $\Delta R_n^{2\rho+1}$. The width of the sparse matrix in (\ref{MasterEq}) does not increase.

\section{Inter- and Extrapolation of the Surface Scattering Kernel}
\label{Extrapolation}
\begin{figure}
\includegraphics[width=\linewidth]{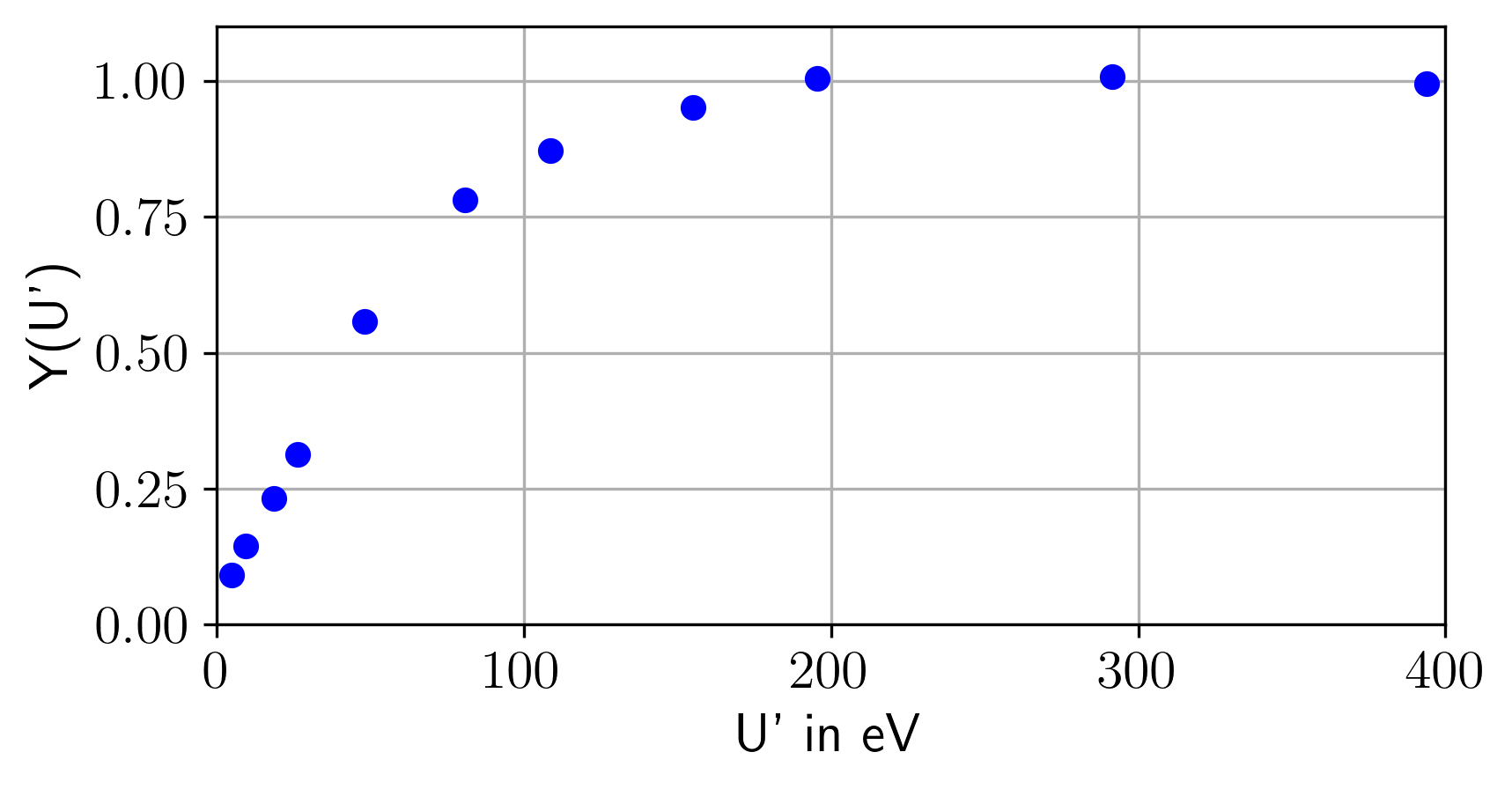}
\caption{The measured value of the EEY \cite{10.1063/1.4984761}, which were used to calculate the correcting factors $\beta_n^{2\rho+1}[j_1|j_2]$.}
\label{EEYmeasured}
\end{figure}
\begin{figure}
\includegraphics[width=\linewidth]{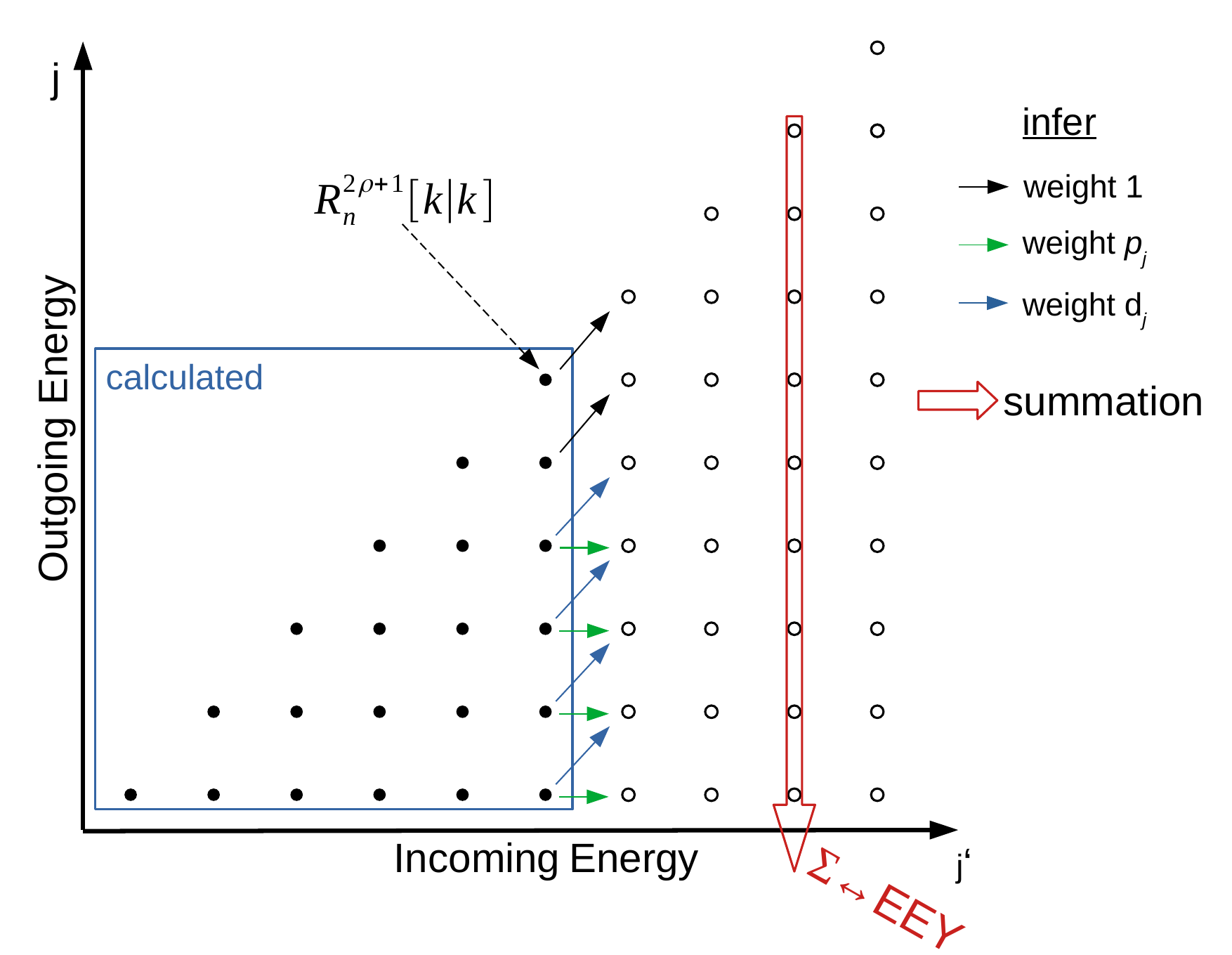}
\caption{The SSK-derivates $R_n^{2\rho+1}(U_j|U_{j'})$ have to be extrapolated. They are already calculated at the solid grid points (in the blue square) up to index $k$. For the details of the algorithm see the main text.}
\label{SketchExtrapol}
\end{figure} 

For a realistic simulation of plasmas, e. g. abnormal glow discharges, which are investigated here, the boundary condition must cover the whole energy range. For the example of Ar, which has a minimal cathode fall voltage of about 150 eV \cite{steenbeckengel}, it becomes clear that the SSK, which is only available up to energies of about 30 eV, needs to be extrapolated to higher energies.

This extrapolation could be done straightforwardly by fitting the computed data at low energies to a sensible function and using this function as the SSK for higher energies \cite{PhysRevSTAB.5.124404}. This method provides a good approximation, if the guessed function fits the data well. Because the extrapolated space is much larger than the actually calculated space, another method is used, which uses measured data for the electron emission yield (EEY) as additional information at higher energies, to get a good approximation of the magnitude of the SSK at higher energies. 

The EEY $Y(U',x')$ can be expressed with the SSK as
\begin{equation}
Y(U',x') = \int\limits_0^{U'} \d U \int\limits_{-1}^0 \d x ~ R(U',x'|U,x)
\end{equation} 
and gives the number of backscattered electrons at any energy and at any angle, when an electron with energy $U'$ and direction cosine $x'$ impinges on a wall. Since measured data for the EEY are rare and most of the time only available for normal incidence \cite{Bronstein,KUHR1999,Joy1995}, the extrapolation is based on the EEY for $x'=1$. The EEY is an integrated quantity of the SSK, hence only the magnitude of the SSK and of the derived $R_n^{2\rho+1}$ can be inferred by the measured EEY, the angular dependencies have to stem from the calculated SSK itself.

To find a usable mathematical link between the EEY and the SSK an additional assumption has to be made, namely that the SSK can be factorized:
\begin{equation}
R(U',x'|U,x) = R_\mathrm{e}(U'|U)\cdot R_\mathrm{a}(x'|x).
\end{equation}
This assumption is also made for some phenomenological models \cite{PhysRevSTAB.5.124404,Dionne1973}, whereas other models \cite{Horvath_2017} use functions that use minor modifications to factorized expressions. With this factorization, the EEY reads
\begin{align}
Y(U',1) &= \int\limits_0^{U'} \d U \int\limits_{-1}^0 \d x ~ R_\mathrm{e}(U'|U)\cdot R_\mathrm{a}(1|x)\nonumber\\
&= \int\limits_0^{U'}\d U ~ R_\mathrm{e}(U'|U)\cdot \int\limits_{-1}^0 \d x ~ R_\mathrm{a}(1|x)\nonumber\\
&= \int\limits_0^{U'}\d U ~ R_\mathrm{e}(U'|U)\cdot \alpha .
\end{align}
Likewise, we can calculate
\begin{align}
\Sigma_n^{2\rho+1}(U') &= \int\limits_0^{U'} \d U ~ R_n^{2\rho+1}(U'|U)\nonumber\\ 
&= \int\limits_0^{U'} \d U ~ R_\mathrm{e}(U'|U) \nonumber\\ 
&~\times \int\limits_{-1}^0 \d x \int\limits_{0}^1 \d x' ~R_\mathrm{a}(x'|x) x^{2\rho+1} P_n(x')\nonumber\\
&= \int\limits_0^{U'}\d U ~ R_\mathrm{e}(U'|U)\cdot \alpha_n^{2\rho+1}.
\label{SigmaInt}
\end{align}
From this one can immediately see that
\begin{equation}
\dfrac{Y(U',1)}{\Sigma_n^{2\rho+1}(U')} = \dfrac{\alpha}{\alpha_n^{2\rho+1}} = \mathrm{const.}
\label{EEY-SSK}
\end{equation}
holds. This relation is the basis for connecting the SSK with the measured EEY presented in Fig. \ref{EEYmeasured}. 

The grid, on which the extrapolation has to be done, is sketched in Fig. \ref{SketchExtrapol}. The $R_n^{2\rho+1}(U_{j^{\,\prime}}|U_j) =: R_n^{2\rho+1}[j^{\,\prime}|j]$ are defined on the grid points and are calculated up to energy $U_k$. To obtain the next column of points, the values of the last calculated column are taken to infer a first guess for the next column, while the angle dependencies are largely maintained and then the sum of all these guesses is compared with the measured EEY and the magnitude of all guessed values is corrected accordingly to fulfill (\ref{EEY-SSK}).

To make a reasonable guess for the next column, one has to distinguish between points on or close to the diagonal and the remaining points, where a significant energy transfer is taking place.  The former ones are in this case the highest and second highest point in each column, because they describe elastic backscattering and the gap in the emission spectrum due to minimum energy transfers in the solid, respectively. These points can only be estimated by points with the same constraints. Thus, the initial guess for these points is simply the value of the corresponding points in the previous column. For the latter points a variety of processes contribute to the backscattering at those index combinations. Therefore, the guesses for those points can make use of multiple values from the previous column. The eligible points from the previous column are those where the outgoing energy $U$ stays constant (leading to parallel inference) and those where the energy transfer $U-U'$ stays constant (leading to diagonal inference).

The two possible contributions to the grid point $[k+1|j]$ must be weighted by the coefficients $p_j$ for parallel inference and $d_j$ for diagonal inference and $p_j+d_j =1$ must hold. Additionally, it is convenient to request $d_0=0$ and $p_{k-1}=0$ to avoid inference from unavailable grid points. This suggests $d_j$ to be a function of $j/(k-1)$. Hence, a new free parameter $\gamma$ is introduced and 
\begin{equation}
d_j = \left( \dfrac{j}{k-1} \right)^\gamma \; \Longleftrightarrow \; p_j = 1-\left( \dfrac{j}{k-1}\right)^\gamma
\end{equation}
is defined. The parameter $\gamma$ needs to be determined by either the calculated or experimental data. A fitting procedure where the second to last calculated column was used to reproduce the last calculated column with the algorithm presented below did not lead to a robust estimate for $\gamma$, therefore experimental spectra were used. Spectra for different materials showed that for an incident energy of 300 eV the EEY decays to 10 \% of its maximal value for an energy of $E_\mathrm{10 \%}\approx 20$ eV. This behaviour was well reproducible with a value of $\gamma = 1.5$, which was used in all simulations. 

For the correction step, (\ref{SigmaInt}) is discretized as
\begin{equation}
\Sigma_n^{2\rho+1}[j^{\,\prime}] = \sum\limits_{j=0}^{j^{\,\prime}} R_n^{2\rho+1}[j^{\,\prime}|j] \cdot \Delta U
\end{equation}
and one can define
\begin{equation}
\beta_n^{2\rho+1}[j_1,j_2] = \dfrac{Y(U_{j_1},1)\cdot \Sigma_n^{2\rho+1}[j_2]}{Y(U_{j_2},1)\cdot \Sigma_n^{2\rho+1}[j_1]}
\end{equation}
which will act as the correcting factor.

The complete algorithm for calculating the values at incoming energy $U_{k+1}$ for the indices $(\rho,n)$ reads:
\begin{enumerate}
\item Set $R_n^{2\rho+1}[k+1|k+1]=R_n^{2\rho+1}[k|k]$ and \\$R_n^{2\rho+1}[k+1|k]=R_n^{2\rho+1}[k|k-1]$
\item Set $R_n^{2\rho+1}[k+1|j]= p_j \cdot R_n^{2\rho+1}[k|j] \\ + d_j\cdot R_n^{2\rho+1}[k|j-1]$ $\forall j<k$
\item Calculate with these estimates $\Sigma_n^{2\rho+1}[k]$,\\ $\Sigma_n^{2\rho+1}[k+1]$ and $\beta_n^{2\rho+1}[k+1,k]$
\item Correct the guesses by setting $R_n^{2\rho+1}[k+1|j] \rightarrow \beta_n^{2\rho+1}[k+1,k] \cdot R_n^{2\rho+1}[k+1|j]$ $\forall j$
\end{enumerate}
This algorithm is then repeated for all $(\rho,n)$ and for all higher energies. 

\begin{figure}
\includegraphics[width=\linewidth]{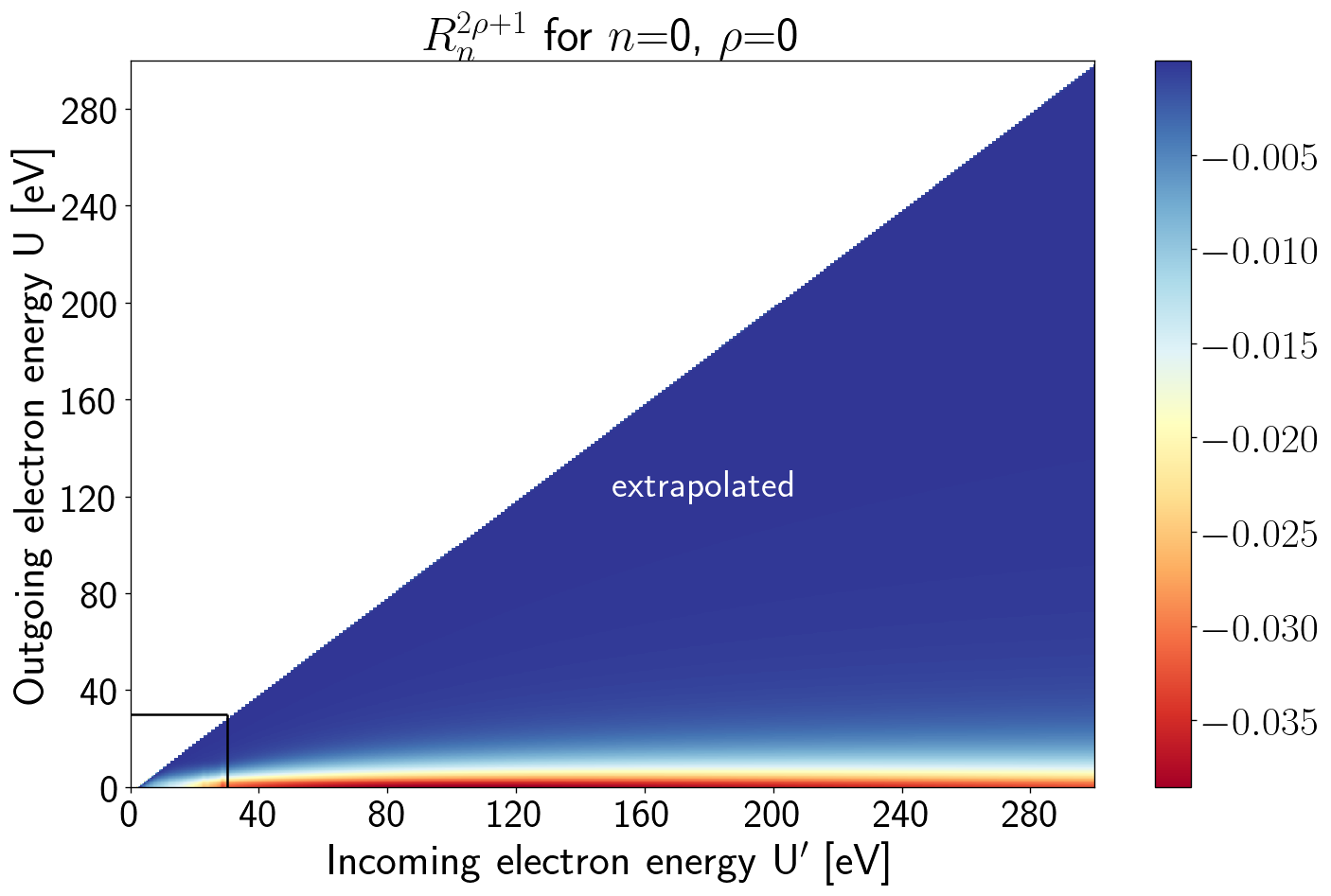}
\caption{$R_0^1$ is shown for energies up to 300 eV. The values inside the black square are calculated. There are no values above the diagonal, because the electrons cannot gain energy inside the wall. The dominance of the emission at low energies is clearly visible.}
\label{fig:SSKExtra}
\end{figure}

The extrapolated values can be seen in Fig. \ref{fig:SSKExtra} for the example of $R_0^1[j|j']$. On one hand, one can see that the SSK is dominated by inelastic backscattering to low energies, which originates from the impact ionization. The energy range to which the secondary electrons are excited does not change linearly with the primary energy but almost becomes constant for high energies. This is due to the extrapolation method, where the values at low backsckattering energy are predominantly inferred by values at the same outgoing energy, and it is reasoned by the impact ionization process itself and noticing that the higher the primary energy is, the lower the secondary electron can have been in the valence band. Thus, the energy of the excited secondary electron after the impact ionization is not so strongly dependent on the incoming electron's energy. 

The dominance of the impact ionization in the relaxation process is also the reason for transferring the angular dependence from the lower energy to the higher energies and not introducing an artificial way of increasing the isotropy. The energy relaxation is almost completely realized by a single scattering event, so the number of scattering events (and with it the isotropy of the scattering cascade) is not proportional to the amount of energy lost.

On the other hand, one can still see that elastic backscattering plays a major role, but not enough to justify an elastic approximation (see Fig. \ref{fig:SSK}).

The energy grid for the plasma simulation is finer than the grid on which the SSK is calculated, hence an interpolation between the grid points for the $R_n^{2\rho+1}[j|j']$ is needed. Here, a simple bilinear interpolation was used, but with respect to the minimum energy transfer in the solid and a strict separation between elastic and inelastic backscattering.\\

\begin{acknowledgments}
We are grateful to D. Loffhagen and M. Becker for valuable discussions. This work was funded by Deutsche Forschungsgemeinschaft (DFG, German Research Foundation) through grant 495729137. 
\end{acknowledgments}


\bibliography{lit.bib}

\end{document}